\documentclass[aps,notitlepage,onecolumn,showpacs,showkeys,preprintnumbers]{revtex4}

\usepackage{graphicx}
\usepackage{amsmath}
\usepackage{amssymb}
\usepackage{bm}
\usepackage{hyperref}

\renewcommand{\rho}{\varrho}

\begin{document}
	
	\title{An explicit numerical scheme for Milne's phase-amplitude equations}
	\author{R. Piron$^{1,2}\footnote{Corresponding author, e-mail address: robin.piron@cea.fr}$ and M. Tacu$^{1,2}$}
	
	\affiliation{$^1$CEA, DAM, DIF, F-91297 Arpajon, France.}
	\affiliation{$^2$Universit\'{e} Paris-Saclay, CEA, Laboratoire Mati\`{e}re en Conditions Extr\^{e}mes, F-91680 Bruy\`{e}res-le-Ch\^{a}tel, France.}
	
	\date{\today}
	
\begin{abstract}
We propose an explicit numerical method to solve Milne's phase-amplitude equations.  Previously proposed methods solve directly Milne's nonlinear equation for the amplitude. For that reason, they exhibit high sensitivity to errors and are prone to instability through the growth of a spurious, rapidly varying component of the amplitude. This makes the systematic use of these methods difficult. On the contrary, the present method is based on solving a linear third-order equation which is equivalent to the nonlinear amplitude equation. This linear equation was derived by Kiyokawa, who used it to obtain analytical results on Coulomb wavefunctions \cite{Kiyokawa15}. The present method uses this linear equation for numerical computation, thus resolving the problem of the growth of a rapidly varying component.
%This method allows to directly . Unlike 
\end{abstract}
	
	%\pacs{xxxx}
	%\keywords{xxxx}
	
\maketitle
	
\section{Introduction}
In atomic physics, estimating the cross sections of collisional processes, photoionization or Bremstrahlung requires the computation of the radial wavefunctions of continuum orbitals. Moreover, in atomic physics of dense plasma, one should in principle account self consistently for continuum electrons in the calculation of the atomic structure (see, e.g., \cite{Blenski13,Piron24} for discussions on this subject). In principle, accounting for the continuum requires sampling the radial wavefunctions in both the radius and momentum spaces (see, for instance, \cite{Liberman79,Wilson06,Piron11,Piron18}). Continuum radial wavefunctions being oscillatory, the sampling limit stated by the Nyquist-Shannon theorem applies \cite{Shannon49}. The minimal grid step for the radius is determined by the largest momentum considered, and the minimal step of the momentum grid is determined by the largest radius considered. Such sampling constraint can quickly lead to computations which are rather intensive in terms of floating-point operations and memory storage, and often constitutes a bottleneck of model implementations. This issue explains, for instance, why a self-consistent, quantum accounting for the continuum is rarely performed in detailed atomic modeling of plasmas.

Alternative formalisms for the one-electron basis may yield a more efficient numerical representation of the continuum. For instance, an average-atom numerical algorithm resorting to the basis of Siegert states was recently proposed and studied \cite{Starrett23}, using the method of solution of \cite{Batishev07}.

However, keeping the usual 1-electron basis, Milne's phase-amplitude representation \cite{Milne30} also enables a more efficient numerical representation of the continuum. It is based on slowly varying, nonperiodic, amplitude and phase functions. Consequently, it has a strong numerical interest, for it allows one to circumvent the Nyquist-Shannon sampling limit. This interest has long been recognized in the literature, and Milne's phase-amplitude representation was used for Bessel functions calculations \cite{Goldstein58} as well as atomic-physics computations \cite{BarShalom96,Wilson03,Wilson06}. Methods of solutions of various kinds were proposed: direct iterative methods were proposed in \cite{Seaton62,Trafton71}, as well as an iterative spectral method using a Chebyshev expansion \cite{Rawitscher15}. A fully explicit, predictor-corrector method resorting to a fixed grid was proposed in \cite{BarShalom96} and used, for instance in the SCROLL \cite{BarShalom97} and HULLAC numerical codes \cite{BarShalom01}. An adaptive mesh refinement Rosenbrock-type method was proposed \cite{Wilson03} and used in the PURGATORIO numerical code \cite{Wilson06}. 

We found that existing methods usually do not allow a simple and systematic use on a pre-defined grid, with a simple relationship between grid coarseness and precision. In particular, they show a rather high sensitivity to numerical errors. In this article, we first summarize the basics of Milne's phase-amplitude representation. We discuss the limitations of some known methods for solving Milne's phase-amplitude equations \cite{BarShalom96,Wilson03,Wilson06}. We then propose a fully explicit method of solution resorting to a fixed grid, which addresses these limitations. Finally, we illustrate the use of this method on some examples.

\section{Phase Amplitude representations}
In the following we will limit the discussion to the description of continuum radial wavefunctions $R(r)$ solution of the radial Schrödinger equation:
\begin{align}
&R''(r)+\omega_{k,\ell}(r)R(r)=0\label{eq_Schrodinger}\\
&\omega_{k,\ell}(r)=k^2-2v(r)-\frac{\ell(\ell+1)}{r^2}
\end{align}
with $k>0$ and $\lim_{r\rightarrow \infty}v(r)=0$.
Most phase-amplitude (PA) representation may also be extended to bound states, considering imaginary $k$. Moreover, PA representations can also be used in the framework of the Dirac radial equation set without any difficulty (see, for instance, \cite{PironPhD}, Section 2.3). 

Phase-amplitude representation refers to a whole category, rather to a particular representation. Such representations consist in recasting the oscillatory wavefunction $R$, regular at zero, as:
\begin{align}
R(r)=A(r)\cdot S\left(\phi(r)\right)
\end{align}
where $S$ is a given oscillatory function regular at zero. $A(r)$ is called the amplitude function, and $\phi(r)$ the phase function.

Generalized PA representation may resort to any oscillatory function $S$ corresponding to the regular solution of the Schrödinger equation for some reference system. The generalized WKB approach \cite{Good53,Rosen64,Wald74,Stein87} may be considered from this standpoint.

In the present article, we are concerned by Milne's PA representation, which uses a sine function for $S$. 
\begin{align}
R(r)=A(r)\cdot \sin\left(\phi(r)\right)\label{eq_RAsinPhi}
\end{align}
The choice of a particular function $S$ is not sufficient to completely determine the representation. In particular, the choice of the sine function for $S$ is used in both Calogero's and Milne's PA representations.

Just to emphasize their differences, let us briefly recall the basics of Calogero's PA representation \cite{Calogero63}. To obtain this PA representation, one requires, in addition to Eq.~\eqref{eq_RAsinPhi}:
\begin{align}
R'(r)=A(r)\cdot \cos\left(\phi(r)\right)\label{eq_Calogero_choice}
\end{align}
So that:
\begin{align}
&A(r)=\sqrt{R(r)^2+R'(r)^2}\\
&\tan\left(\phi(r)\right)=\frac{R(r)}{R'(r)}
\end{align}
$R(r)$ being regular at zero, Calogero's amplitude function is regular at zero.
Re-writing Eq.~\eqref{eq_Schrodinger}, using Eqs~(\ref{eq_RAsinPhi}, \ref{eq_Calogero_choice}) yields Calogero's equations for the phase and amplitude (see \cite{Calogero63}, Eqs~(6.16) and (4.7)):
\begin{align}
\phi'(r)=k+\frac{1}{k}\left(k^2-\omega_{k,\ell}(r)\right)\sin^2\left(\phi(r)\right)\\
A'(r)=A(r)\left(k^2-\omega_{k,\ell}(r)\right)\frac{\sin\left(2\phi(r)\right)}{2k}
\end{align}

From these equations, it appears clearly that the amplitude and phase functions of Calogero's PA representation exhibit rapid variations, i.e. on the scale of $r\sim 1/(2k)$. This representation was used to derive many useful analytical results, as for instance an analytic approach to phase shifts in screened potentials \cite{Green82}. Calogero's PA representation was also used in order to build efficient numerical methods for searching the eigenvalues of bound states in an attractive potential \cite{Blenski88, Nikiforov}. However, when applied to continuum orbitals, this representation requires the same typical sampling of $A(r)$, $\phi(r)$ as the radial wavefunction itself.

Milne's PA representation is most often derived by inserting Eq.~\eqref{eq_RAsinPhi} into Eq.~\eqref{eq_Schrodinger}, yielding
\begin{align}
 \left[
2A'(r)\phi'(r)+A(r)\phi''(r)
\right]\cdot\cos\left(\phi(r)\right)
+\left[
A''(r)-A(r)\left(\phi'(r)\right)^2+\omega_{k,\ell}(r)A(r)
\right]\cdot\sin\left(\phi(r)\right)
=0
\end{align}
One then requires the factors in front of the sine and cosine of the phase to be zero independently. This leads to Milne's equations for the phase and amplitude, respectively:
\begin{align}
&\phi'(r)=\frac{\theta}{A(r)^2}\label{MPA_eq_phase}\\
&A''(r)+\omega_{k,\ell}(r)A(r)-\frac{\theta^2}{A(r)^3}=0
\label{MPA_eq_amp}
\end{align}
with $\theta$ appearing as an arbitrary integration constant.

A somewhat more explicit definition of Milne's phase and amplitude may be stated as follows. Let us consider $R(r)$, $Q(r)$ two linearly independent solutions of Eq.~\eqref{eq_Schrodinger}, with $R(r)$ being regular at $r=0$, and define $A(r)$, $\phi(r)$ as follows:
\begin{align}
R(r)=A(r)\cdot \sin\left(\phi(r)\right)\\
Q(r)=A(r)\cdot \cos\left(\phi(r)\right)
\end{align}
One has
\begin{align}
&A(r)=\sqrt{R(r)^2+Q(r)^2}\label{MPA_def_A}\\
&\tan\left(\phi(r)\right)=\frac{R(r)}{Q(r)}\label{MPA_def_Phi}
\end{align}
Differentiating Eq.~\eqref{MPA_def_Phi}, one gets:
\begin{align}
\left(1+\tan^2\left(\phi(r)\right)\right)\phi'(r)&=
\frac{Q(r)R'(r)-R(r)Q'(r)}{Q(r)^2}\\
\phi'(r)&=\frac{\mathcal{W}\{Q(r),R(r)\}}{A(r)^2}\label{MPA_eq_phase_2}
\end{align}
where $\mathcal{W}$ denotes the Wronskian.
Differentiating Eq.~\eqref{MPA_def_A} two times, one gets:
\begin{align}
A''(r)=&\frac{-1}{A(r)^2}A'(r)\left[R(r)R'(r)+Q(r)Q'(r)\right]
+\frac{1}{A(r)}\left[R(r)R''(r)+R'(r)^2+Q(r)Q''(r)+Q'(r)^2\right]
\end{align}
using Eq.~\eqref{eq_Schrodinger} for $R(r)$ and $Q(r)$, and the definition of Eq.~\eqref{MPA_def_A}, one gets:
\begin{align}
A''(r)=&-\omega_{k,\ell}(r)A(r)-\frac{1}{A(r)}
\left[A'(r)^2-R'(r)^2-Q'(r)^2\right]
=-\omega_{k,\ell}(r)A(r)+\frac{\mathcal{W}\{Q(r),R(r)\}^2}{A(r)^3}
\label{MPA_eq_amp_2}
\end{align}
Defining $\theta=\mathcal{W}\{Q(r),R(r)\}$, Eqs~\eqref{MPA_eq_phase_2} and \eqref{MPA_eq_amp_2} become Eqs~\eqref{MPA_eq_phase} and \eqref{MPA_eq_amp}, respectively. This shows how $\theta$ is related to the normalizations of the $R$ and $Q$ functions.

Two particular cases are those of free particles ($v(r)=0$) and particles in a Coulomb potential ($v(r)=-Z/r$). For these, one has the analytic solutions:
\begin{align}
R(r)=
\begin{cases}
kr\,j_\ell(kr)&\text{($v(r)=0$)}\\
F_{\ell} (\eta=-Z/k,kr)&\text{($v(r)=-Z/r$)}
\end{cases}
\label{eq_R_Bessel_Coulomb}
\\
Q(r)=
\begin{cases}
-kr\,y_\ell(kr)&\text{($v(r)=0$)}\\
G_\ell (\eta=-Z/k,kr)&\text{($v(r)=-Z/r$)}
\end{cases}
\label{eq_Q_Bessel_Coulomb}
\end{align}
with $j_\ell$, $y_\ell$ being the regular and irregular sperical Bessel functions, respectively, and $F_\ell$, $G_\ell$ being the regular and irregular Coulomb wavefunctions. We use the conventions of  \cite{AbramowitzStegun}. The latter analytic solutions can be used, together with Eq.~\eqref{MPA_def_A} for the evaluation of the boundary condition for fully screened and Coulomb-tail atomic potentials, respectively.

Equation~\eqref{MPA_eq_amp} is a second-order nonlinear differential equation. Its solutions are nontrivial. However, in the asymptotic limit $r\rightarrow\infty$, Eq.~\eqref{MPA_eq_amp} simply becomes:
\begin{align}
&A''(r)+k^2 A(r)-\frac{\theta^2}{A(r)^3}=0
\label{MPA_eq_amp_asympt}
\end{align}
One readily see that a constant $A(r)$ can be solution of this equation, its value being:
\begin{align}
&A_\text{slow}=\left(\frac{\theta}{k}\right)^{1/2}
\label{MPA_eq_amp_asympt_cst_sol}
\end{align}
which directly yields $\phi_\text{slow}(r)\sim kr$ using Eq.~\eqref{MPA_eq_phase}.

This asymptotic limit may also be recovered from the usual WKB approximation, which just consists in neglecting the second derivative $A''(r)$ in Eq.~\eqref{MPA_eq_amp}:
\begin{align}
&A_\text{WKB}(r)=\frac{\theta^{1/2}}{\omega_{k,\ell}(r)^{1/4}}
\xrightarrow[r \to \infty]{}
A_\text{slow}\\
&\phi_\text{WKB}'(r)=\omega_{k,\ell}(r)^{1/2}\xrightarrow[r \to \infty]{}
k
\end{align}

Let us stress out that the parameter $\theta$ is directly related through Eq.~\eqref{MPA_eq_amp_asympt_cst_sol} to the asymptotic value of the amplitude function, which determines the normalization of $R(r)$. $\theta$ therefore has to be consistent with the boundary condition applied when solving Eq.~\eqref{MPA_eq_amp} for the amplitude. Having a parameter in the equation which has to be consistent with the boundary condition is typical for a nonlinear equation. On the contrary, linear equations allow the boundary condition to be set with an arbitrary multiplicative constant.

The solution of Eq.~\eqref{MPA_eq_amp} fulfilling the condition $\lim_{r\rightarrow\infty}A''(r)=0$ corresponds to a slowly varying amplitude function, asymptotically tending to the value $A_\text{slow}$. This is the amplitude function we are interested in, for it allows us to circumvent the Nyquist-Shannon sampling limit related to the oscillatory wavefunction $R(r)$.
% This has motivated the search for numerical methods of computing the slowly varying amplitude function.

However, Eq.~\eqref{MPA_eq_amp} may have other solutions than the slowly varying one. Going back to its asymptotic limit Eq.~\eqref{MPA_eq_amp_asympt}, let us try to find another solution by considering a first-order perturbation $A_\text{slow}+\delta A(r)$ around the constant solution of Eq.~\eqref{MPA_eq_amp_asympt_cst_sol}. The equation for the perturbation is:
\begin{align}
&\delta A''(r)+k^2 \delta A(r)+3\frac{\theta^2}{A_\text{slow}^4}\delta A(r)=0\\
&\delta A''(r)=-4k^2 \delta A(r)
\end{align}
From which we get $\delta A(r)\sim e^{2ikr}$, showing that other solutions exist, and that they exhibit rapid variations, over the $1/(2k)$ scale.

Equation~\eqref{MPA_eq_amp} is thus a stiff differential equation in the sense that it has multiple solutions varying over very different scales. However, for the purpose of sampling reduction, we are only interested in the slowly varying one. Moreover, being nonlinear, this equation can allow coupling among its various solutions, which may constitute a problem, as we will see in next section. 

\begin{figure}[t]
\centerline{
\includegraphics[width=8cm]{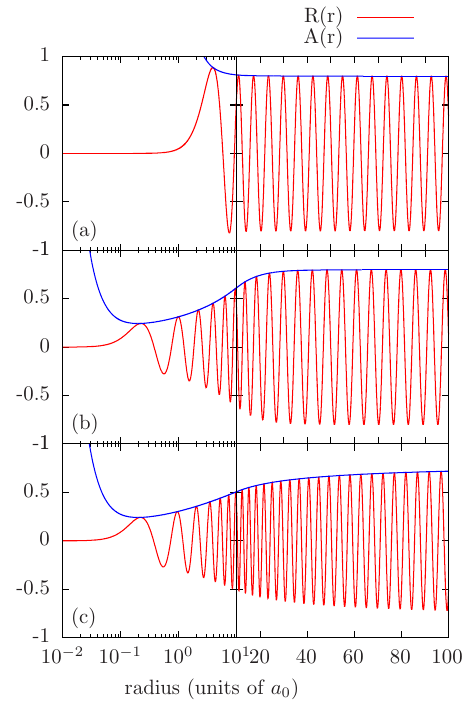}
}
\caption{Radial wavefunction and amplitude function $\ell=2$, $k=1$ for three distinct potentials. Zero potential (a), Yukawa potential $-Z e^{-r/\lambda}/r$ with $Z=26$, $\lambda=10$ (b), and pure Coulomb potential $-Z/r$ with $Z=26$ (c).
\label{fig_illustration_amp}}
\end{figure}	

For the sake of illustrating the slowly varying amplitude function of Milne's PA representation, we present in Fig.~\ref{fig_illustration_amp} the radial wavefunctions $R(r)$ and the corresponding amplitude functions $A(r)$, for $k=1\,a_0^{-1}$, $\ell=2$, in the case of a zero potential (a), of a screened Coulomb potential (b), and of a pure Coulomb potential (c). The slowly varying amplitude function give the enveloppe of the wavefunction $R(r)$.

Of course, the amplitude function can be obtained from a numerical solution of Eq.~\eqref{eq_Schrodinger}, finding the two linearly independent solutions $R$ and $Q$ and then combining them as in Eq.~\eqref{MPA_def_A}. However, such a method would require sampling oscillatory functions, thus defeating the purpose of using the PA representation.

For that reason a number of authors have studied numerical methods for solving Eq.~\eqref{MPA_eq_amp} directly. Seaton and Peach, as well as Trafton proposed iterative schemes, starting from an initial guess of the amplitude \cite{Seaton62,Trafton71}. Bar Shalom, Klapisch and Oreg  proposed in \cite{BarShalom96} to use of a predictor-corrector due to Hamming (see \cite{Hamming} \textsection 24.3) when it is stable, and proposed a modified predictor-corrector for the region in which the latter method is unstable. In \cite{Wilson03, Wilson06}, Wilson, Sonnad, Sterne and Isaacs proposed to use a Kaps-Rentrop method \cite{Kaps79,NumericalRecipes}. In \cite{Rawitscher15} Rawitscher proposed a spectral iterative method based on a Chebyshev expansion.

In atomic-physics applications, description of the continuum usually requires the calculation of a large number of wavefunctions, in order to sample the momentum space. For that reason, there is a special interest in fast, fully explicit numerical methods, such as that proposed in \cite{BarShalom96}. Before coming to the method that we propose, let us first motivate the present work by discussing the limitation of some existing methods.

\section{Limitations of some existing methods}

In \cite{BarShalom96}, Bar Shalom et al. propose an explicit method to solve numerically Eq.~\eqref{MPA_eq_amp} on a fixed grid of exponentially spaced points. The method in fact resorts to three different numerical schemes. Let us recall briefly its general principle, because it guided us on many aspects in building our method.

In the innermost region, a standard Numerov scheme is used to solve Eq.~\eqref{eq_Schrodinger} for $R(r)$, up to a matching point $r_\text{match}$, chosen close to the turning point $r_\text{turn}$, such that $\omega_{k,\ell}(r_\text{turn})=0$. 

Beyond the matching point, Bar Shalom et al. use a predictor-corrector method due to Hamming (see \cite{Hamming} \textsection 24.3), in order to solve Eq.~\eqref{MPA_eq_amp} for $A(r)$. Let us call this scheme Hamming's predictor-corrector (HPC).
However, they find that, as soon as $h^2\omega_{k,\ell}(r)/3$ is greater than one, then the corrector step amplifies the prediction error.
Far from the origin, the slowly varying amplitude solution tends to a constant. $A''(r)$ thus takes small values, which are obtained through Eq.~\eqref{MPA_eq_amp} as differences between two values that are comparatively large. This leads to large numerical errors. This issue is in fact common to all approaches that make use of Eq.~\eqref{MPA_eq_amp} to obtain $A''(r)$ from $A(r)$.

As a solution to this problem,  Bar Shalom et al. propose a modified predictor-corrector scheme for the outermost region. Its basic principle is to use the fact that $A(r)$ is varying slowly to build a predictor for $A''(r)$ and then use Eq.~\eqref{MPA_eq_amp} as a corrector, deducing a new value of $A(r)$ from $A''(r)$, performing one step of a Newton method. Using the amplitude equation in this other way, the error amplification factor is inverted, leading to a stable scheme. Let us call this scheme Bar Shalom modified predictor corrector (BSMPC).

In addition to the BSMPC scheme, the authors suggest to further improve the solution in the outermost region by re-evaluating $A''(r)$ from a high order finite-difference scheme (5-point scheme) and correcting again using Eq.~\eqref{MPA_eq_amp}. Adding such a step makes the method become partly an iterative method and we did not find it useful to resort to such improvement in the present study.

The switching radius $r_\text{switch}$ from HPC to BSMPC scheme is located where $h^2\omega_{k,\ell}(r)/3\approx 1$ and these two schemes are actually employed in order to propagate the solution inwards, starting from the outer boundary condition at $r_\text{max}$. Finally, solutions for $R(r)$ and $A(r)$ are matched at the radius $r_\text{match}$, allowing one to normalize $R(r)$ and determine the phase at the matching point.
A table summarizing this method is given in Table~\ref{tab_regions_BS}

\begin{table}[b]
\begin{tabular}{c|c|c}
Innermost region &	Intermediate region 		&  Outermost region \\
$[r_\text{min},r_\text{match}]$ & $[r_\text{match},r_\text{switch}]$ & $[r_\text{switch}, r_\text{max}]$ \\
\hline
$h^2\omega_{k,\ell}(r)/3<1$ & $h^2\omega_{k,\ell}(r)/3<1$ & $h^2\omega_{k,\ell}(r)/3>1$\\
Numerov scheme for $R(r)$ &
Hamming's predictor corrector&
Bar Shalom's modified predictor-corrector \\
Outwards & Inwards & Inwards
\end{tabular}
\caption{Summary of the numerical method proposed in \cite{BarShalom96}.\label{tab_regions_BS}}
\end{table}

\begin{figure}[t]
\centerline{
\includegraphics[width=8cm]{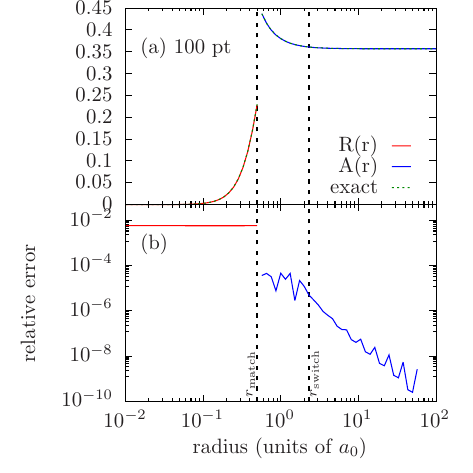}
\includegraphics[width=8cm]{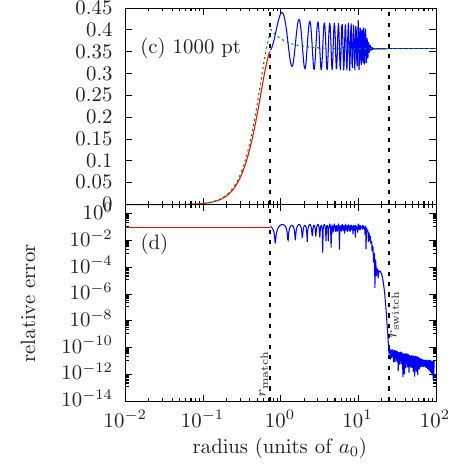}
}
\caption{Radial wavefunction and amplitude function (a,c), with the corresponding relative errors (b,d) obtained with the method of \cite{BarShalom96} for the case of zero potential, $\ell=2$, $k=5\,a_0^{-1}$, using a exponentially spaced grid spanning from $10^{-4}$ to $10^2$\,$a_0$. The reference, exact solution are shown in dashed green. Comparison between grids having $100$ (a,b) and $1000$ (c,d) grid points in total, respectively. The $100$-points grid has $38$ steps between $r_\text{match}$ and $r_\text{max}$, whereas the $1000$-points grid has $272$ steps.
\label{fig_limitation_BarShalom}}
\end{figure}	

We found that the numerical method proposed in \cite{BarShalom96} performs very well on coarse grids. As described in their article, one really needs to apply both HPC and BSMPC schemes, paying attention to the switching radius $r_\text{switch}$. It may be checked that error amplification actually occurs when using any of the numerical scheme on the wrong side of $r_\text{switch}$, and that the overall method is very sensitive to such error amplification. 

In his book, Hamming derives a predictor-corrector numerical scheme of order 6. He also describes corrections to the predictor and corrector (see \cite{Hamming}, end of  \textsection 24.3), which are presented as optional, and makes the scheme of order 8. We found that in the present case, these corrections have a significant impact on the global stability of the numerical method, and assume that Bar Shalom and coauthors did use these corrections.

In principle, if one wants to increase the precision of the solution, refining the grid should be the way to proceed. It turns out that beyond a certain level of refinement, a numerical instability occurs, ruining the numerical solution. This issue is illustrated on Fig.~\ref{fig_limitation_BarShalom}, which displays the results obtained in a case of zero potential (amplitude of spherical Bessel functions), when increasing the number of grid points from 100 to 1000.

The triggering of an instability when considering \emph{finer} grids constitutes a serious limitation to a systematic use of this numerical scheme, as it impairs the control of the numerical precision.

On Fig.~\ref{fig_limitation_BarShalom}\,d, one may clearly identify a region of rapid growth of error. This region seems to always lie in the intermediate region, where HPC scheme is used. By applying step by step the predictor-corrector scheme on a test case for which the solution is known exactly, one may analyze precisely the source of errors. Restarting at each step from the ``exact'' amplitude, one can see that the error growth is driven by the predictor, whereas the corrector systematically improves the solution as it is expected to do.

An explanation for this large error growth may be the nonlinear character of Eq.~\eqref{MPA_eq_amp}. In principle, it allows the coupling among various solutions and enables the growth of a spurious, rapidly varying component in the solution, starting from an initial seed.

In principle, we select the desired, slowly varying solution through the boundary condition. However, any error, either the truncation error on the boundary condition, or the accumulation of numerical errors when propagating the solution, may be seen as a mixing between the desired solution and the spurious ones, seeding the growth of the rapidly varying component.

A numerical limitation to the growth of a rapidly varying component is the sampling. A coarse grid acts as a low-pass filter a may temper the growth of a rapidly varying component. It turns out that the criterion for switching from HPC to BSMPC is close to the Nyquist-Shannon sampling criterion for the rapidly varying solution. This probably explains why the region where the spurious component builds up is located inside the intermediate region, near the switching radius.

Whatever the method used, any attempt to solve directly the nonlinear Eq.~\eqref{MPA_eq_amp} is likely to yield a result which is strongly sensitive to errors and prone to the growth of a rapidly varying component in the amplitude. This notably explains the crucial need for a high-order scheme in the method of \cite{BarShalom96}, and why the corrections which allow one to reach order 8 matter so much.

In \cite{Wilson03,Wilson06}, the authors used a Rosenbrock method proposed by Kaps and Rentrop \cite{Kaps79}, frequently used for stiff equations \cite{NumericalRecipes}. Typical numerical methods for stiff equations are usually based on adaptive grid refinement, aiming at correctly sampling all solutions, with all their various scales of variation. Certainly Eq.~\eqref{MPA_eq_amp} is stiff, but our purpose is only to calculate its slowly varying solution, and not to achieve a correct sampling of all of its solutions. In this view, it is likely that such methods are of limited help for the present problem. However, putting a very stringent precision target, adaptive grid refinement may also limit the accumulation of errors, therefore limiting the seed for a spurious component. One should however keep in mind that, in many cases of application, the potential $v(r)$ associated to the wavefunctions is sampled on a predefined grid. Adaptive refinement methods then require to interpolate the potential at intermediate grid points, resulting in a supplementary loss of precision.

\begin{figure}[t]
\centerline{
\includegraphics[width=8cm]{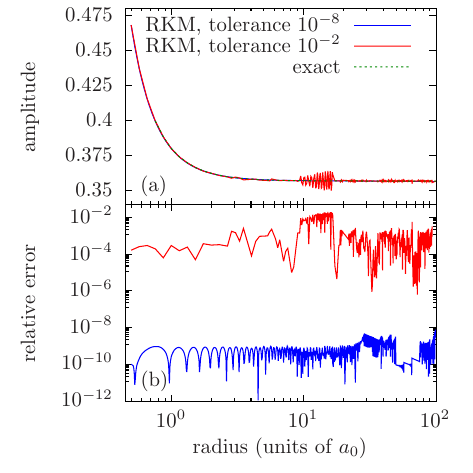}
}
\caption{Amplitude function (a), with the corresponding relative error (b) obtained with the Runge-Kutta-Merson adaptive refinement method, for the case of zero potential, $\ell=2$, $k=5\,a_0^{-1}$, using an initial grid of $100$ exponentially spaced points, spanning from $10^{-4}$ to $10^2$\,$a_0$. Presented results correspond to precision targets (or tolerances) of $10^{-2}$ and $10^{-8}$. In the present case, the refined grids actually have $436$ and $848$ grid points between $r_\text{match}$ and $r_\text{max}$, respectively, to be compared with the $38$ points of the initial grid.
\label{fig_limitation_RKM}}
\end{figure}	

A simple, explicit method resorting to adaptive grid refinement is the Runge-Kutta-Merson (RKM) method \cite{Merson57}. We choose this method as a test-bed in order to illustrate a typical issue of using adaptive grid refinement method for Eq.~\eqref{MPA_eq_amp}. 

The RKM method is based on a 4$^\text{th}$ order Runge-Kutta (RK4) scheme, with a supplementary evaluation which allows one to estimate the truncation error. Let us first put Eq.~\eqref{MPA_eq_amp} under the canonical form:
\begin{align}
\begin{cases}
\bar{A}_0'(r)=\bar{A}_1(r)&\equiv \bar{f}_0(r;\bar{A})\\
\bar{A}_1'(r)=-\omega_{k,\ell}(r)\bar{A}_0(r)+\dfrac{\theta^2}{\bar{A}_0(r)^3}&\equiv \bar{f}_1(r;\bar{A})
\end{cases}
\Leftrightarrow \bar{A}'(r)=\bar{f}(r;\bar{A})
\end{align}
with $\bar{A}_0(r)$ being $A(r)$, the bars denoting vector functions having two components labelled $0$ and $1$, respectively.

Solving for $\bar{A}$ over one step $h$ with a classical RK4 scheme requires the evaluation of the function $\bar{f}$ at four different points:
\begin{align}
&k_1 = hf(r;\bar{A}) 
& k_3 = hf\left(r+\frac{1}{3}h;\bar{A}+\frac{1}{6}k_1+\frac{1}{6}k_2 \right) \\
&k_2 = hf\left(r+\frac{1}{3}h;\bar{A}+\frac{1}{3}k_1\right)
& k_4 = hf\left(r+\frac{1}{2}h;\bar{A}+\frac{1}{8}k_1+\frac{3}{8}k_3 \right)
\end{align}
One then expresses the fourth-order-approximate solution $\bar{A}(x+h)$ as  
a weighted sum of the $k_i$'s.
\begin{equation}
\bar{A}(r+h) = \bar{A}(r) + \frac{k_1}{2} - \frac{3}{2}k_3 + 2k_4.
\end{equation}
Then, using only one more evaluation of the function $\bar{f}$:
\begin{equation}
k_5 = hf\left(r+h;\bar{A}+\frac{k_1}{2}-\frac{3}{2}k_3 + 2k_4 \right)
\end{equation}
one may calculate a fifth-order approximation $\bar{A}^*(r+h)$:
\begin{equation}
\bar{A}^*(r+h) = \bar{A}(r) + \frac{1}{6}k_1 + \frac{2}{3}k_4+\frac{1}{6}k_5.
\end{equation}
The truncation error $R = 0.2|\bar{A}^* - \bar{A}|$ can then be estimated, allowing one to choose the integration step according to a fixed precision target, or tolerance $\epsilon$. The algorithm that we use is as follows. If $R>\epsilon$, divide the integration step $h$ by $2$. If $R<\epsilon/64$, multiply it by $2$. If $\epsilon/64 < R < \epsilon$, accept the step, and go to the next grid point $r+h$.

As for the method of Bar Shalom et al., we actually employ the RKM scheme to propagate the solution inwards, starting from the outer boundary condition at $r_\text{max}$. We apply the RKM method successively for each interval between two points of the base grid.

Figure~\ref{fig_limitation_RKM} presents the results obtained with the RKM method on the same case as in Fig.~\ref{fig_limitation_BarShalom}, the base grid being that of subfigures a and b ($100$ points, out of which $38$ for the $[r_\text{match},r_\text{max}]$ interval).

Setting the tolerance to $10^{-2}$, one can see how the spurious solution can grow, requiring the method to refine the grid in order to correctly sample its rapid variations. In the present case, the refined grid actually has 436 points, to be compared with the 38 points between $r_\text{match}$ and $r_\text{max}$ of the initial grid.

Decreasing the tolerance of the integration steps to $10^{-8}$, one may efficiently prevent the growth of numerical errors but this is done at the price of  refining the grid even more. In the present case, the refined grid for tolerance $10^{-8}$ has 848 grid points.

\begin{figure}[h]
\centerline{
\includegraphics[width=8cm]{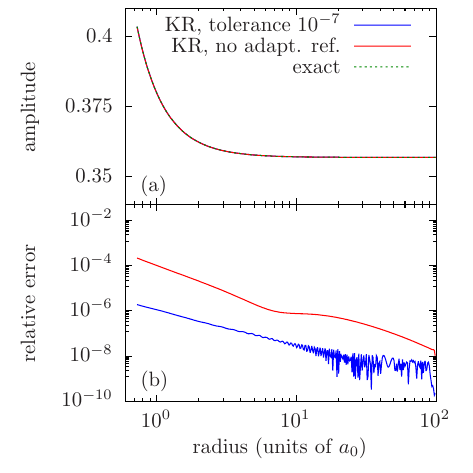}
}
\caption{Amplitude function (a), with the corresponding relative error (b) obtained with the Kaps-Rentrop method, for the case of zero potential, $\ell=2$, $k=5\,a_0^{-1}$, using a exponentially spaced base grid of $1000$ points, spanning from $10^{-4}$ to $10^2$\,$a_0$. Results obtained with and without adaptive grid refinement are presented. In the present case, the refined grid corresponding to precision target of $10^{-7}$ actually has $16221$ grid points between $r_\text{match}$ and $r_\text{max}$, to be compared with the $272$ points of the initial grid.
\label{fig_limitation_KapsRentrop}}
\end{figure}

The Kaps-Rentrop (KR) method \cite{Kaps79,NumericalRecipes} was used by Wilson et al. to solve the amplitude equation related to Dirac equation, which is an analogous of Eq.~\eqref{MPA_eq_amp} \cite{Wilson03,Wilson06}. It is an adaptive mesh refinement method based on an implicit scheme.

Figure~\ref{fig_limitation_KapsRentrop} presents the results obtained with KR method on the same case as in Fig.~\ref{fig_limitation_BarShalom}, the base grid being that of subfigures c and d ($1000$ points, out of which $272$ for the $[r_\text{match},r_\text{max}]$ interval). We tried using the KR numerical scheme both with and without performing an adaptive grid refinement.

It may be seen on Fig.~\ref{fig_limitation_KapsRentrop}\,a that the implicit scheme of the KR method seems stable over the whole region between $r_\text{match}$ and $r_\text{max}$. Using directly the base grid, that is, performing no adaptive refinement, the KR numerical scheme yields a result of limited precision, with maximum relative error being on this case about $10^{-4}$. In order to improve precision, grid refinement has to be performed. In order to achieve a maximum relative error about $10^{-6}$, the number of effective grid points reaches $16221$, to be compared with the $272$ points of the base grid.

Moreover, all these methods are very sensitive to errors on the boundary condition. Any inconsistency between the boundary condition and the value of $\theta$ is viewed as a contribution from a rapidly varying component, which results in heavy grid refinement as soon as its amplitude reaches the precision target (see, for instance, \cite{PironPhD}, Fig.~6.9, pp.~115).

\section{Proposed numerical method}
\subsection{General principle}
In reference \cite{Kiyokawa15}, Kiyokawa uses the following transformation, initially pointed out by Trafton (see \cite{Trafton71}, Eq.~(8,9)). Multiplying Eq.~\eqref{MPA_eq_amp} by $2\,A'(r)$, and rearranging the terms, one can write:
\begin{align}
\frac{d}{dr}\left(\left(A'(r)\right)^2+\frac{\theta^2}{A(r)^2}\right)
+\omega_{k,\ell}(r)\frac{d}{dr}\left(A(r)^2\right)=0\label{eq_Trafton}
\end{align}
and perform the change of function $Y(r)=A(r)^2$. It was noted by Kiyokawa that, if one expresses $Y''(r)$ and uses Eq.~\eqref{MPA_eq_amp}, one obtains the relation:
\begin{align}
\left(A'(r)\right)^2=\frac{Y''(r)}{2}+\omega_{k,\ell}(r)Y(r)-\frac{\theta^2}{Y(r)}
\end{align}
When inserted into Eq.~\eqref{eq_Trafton}, the latter relation yields a \emph{linear} third-order differential equation for $Y$:
\begin{align}
Y^{(3)}(r)+4\omega(r)Y'(r)+2\omega'(r)Y(r)=0\label{eq_amp_lin}
\end{align}
$Y^{(3)}$ denoting the third derivative of $Y$. Kiyokawa used this linear equation in order to derive analytic results on Coulomb wavefunctions \cite{Kiyokawa15}.

We propose in this article to use Eq.~\eqref{eq_amp_lin} for numerical calculations, instead of Eq.~\eqref{MPA_eq_amp}. In principle, using a linear equation should exclude any coupling to spurious, rapidly varying, solutions and completely solve the issue of sensitivity to errors. We should then obtain a numerical method that is much more robust and can be applied more systematically. In particular, it should be tolerant with respect to any truncation error on the boundary condition. In this respect, it may be worth noting that in Eq.~\eqref{eq_amp_lin}, the parameter $\theta$ disappears, and the boundary condition may be defined with any multiplicative constant, as is expected and necessary for a linear equation.

In practice, the numerical sheme we propose is mostly an extension of the well-known Numerov scheme, plus an adaptation of the modified predictor-corrector scheme of Bar Shalom et al. \cite{BarShalom96}.

\subsection{Innermost region}
Milne's amplitude function is singular at zero. For that reason, as is done in \cite{BarShalom96}, we do not solve for the amplitude function close to zero. We instead solve for the wavefunction $R(r)$ using a standard Numerov scheme, propagating the solution outwards. In order to minimize the error when matching, we choose the matching point to be located near the first maximum of the wavefunction, or, if it is closer to zero, to the radius at which a chosen fraction of the sampling limit is reached.

\subsection{Intermediate region}

We recast the linear 3rd-order equation Eq.~\eqref{eq_amp_lin} under the partial canonical form that follows:
\begin{align}
&Y_1''(r)+4\omega(r)Y_1(r)=-2\omega'(r)Y_0(r)\label{eq_amp_lin_cano1}\\
&Y_0'(r)=Y_1(r)\label{eq_amp_lin_cano0}
\end{align}
where $Y_0(r)\equiv Y(r)$. We treat the first of these two equations using a standard Numerov scheme, and the second one using a corollary result of Numerov method for the derivative.

Let us consider the Taylor expansion for the $q$-th derivative of $Y_0$, $Y_1$:
\begin{align}
Y_\alpha^{(q)}(r\pm h)
=&Y_\alpha^{(q)}(r)
\pm h Y_\alpha^{(q+1)}(r) 
+ \frac{h^2}{2}Y_\alpha^{(q+2)}(r)
\pm \frac{h^3}{6} Y_\alpha^{(q+3)}(r) 
+ \frac{h^4}{24} Y_\alpha^{(q+4)}(r)%\nonumber\\
&\pm \frac{h^5}{120} Y_\alpha^{(q+5)}(r) 
+O(h^6)
\label{eq_taylor}
\end{align}
Let us consider that $r$ is uniformly sampled: $r_n=r_\text{min}+n\,h$. Using the Taylor expansion at $r+h$ and $r-h$, we may write:
\begin{align}
Y_{\alpha,n+1}^{(q)}+Y_{\alpha,n-1}^{(q)}
=&2 Y_{\alpha,n}^{(q)}
+ h^2 Y_{\alpha,n}^{(q+2)}
+ \frac{h^4}{12} Y_{\alpha,n}^{(q+4)}
+O(h^6)
\label{eq_taylor_plus}\\
Y_{\alpha,n+1}^{(q)}-Y_{\alpha,n-1}^{(q)}
=&2 h Y_{\alpha,n}^{(q+1)}
+ \frac{h^3}{3} Y_{\alpha,n}^{(q+3)}
+ \frac{h^5}{60} Y_{\alpha,n}^{(q+5)}
+O(h^7)
\label{eq_taylor_minus}
\end{align}
where the $n$ indices denote values of the functions taken at $r_n$.

First, we use Eq.~\eqref{eq_taylor_plus} with $\alpha=1$, $q=0$ and with $\alpha=1$, $q=2$, and eliminate $Y_{\alpha,n}^{(4)}$.
\begin{align}
Y_{1,n+1}+Y_{1,n-1}
=&2 Y_{1,n}
+ h^2 Y''_{1,n}
+ \frac{h^2}{12} \left( Y''_{1,n+1}+Y''_{1,n-1}-2 Y''_{1,n} \right)
+O(h^6)
\end{align}
Using Eq.~\eqref{eq_amp_lin_cano1}, we obtain:
\begin{align}
Y_{1,n+1}\left(1+\frac{h^2}{3}\omega_{n+1}\right)
=Y_{0,n+1}\left(-\frac{h^2}{6}\omega'_{n+1}\right)
+Y_{1,n}\left(2-\frac{10h^2}{3}\omega_{n}\right)
+Y_{0,n}\left(-\frac{10h^2}{6}\omega'_{n}\right)\nonumber\\
+Y_{1,n-1}\left(-1-\frac{h^2}{3}\omega_{n-1}\right)
+Y_{0,n-1}\left(-\frac{h^2}{6}\omega'_{n-1}\right)
+O(h^6)
\label{eq_scheme_Y1_Numerov}
\end{align}
which corresponds to a usual Numerov scheme for a second-order inhomogeneous equation with no first-order derivative.

Then, in order to express $Y_{0,n+1}$, we use Eq.~\eqref{eq_taylor_minus} with $\alpha=0$, $q=0$, and with $\alpha=0$, $q=2$, and eliminate $Y_{0,n}^{(3)}$.
\begin{align}
Y_{0,n+1}-Y_{0,n-1}=2hY'_{0,n}+\frac{h^2}{6}\left( Y''_{0,n+1}-Y''_{0,n-1} \right)+O(h^5)
\end{align}
Using Eq.~\eqref{eq_amp_lin_cano0} and its derivative, we obtain.
\begin{align}
Y_{0,n+1}-Y_{0,n-1}=2hY_{1,n}+\frac{h^2}{6}\left( Y'_{1,n+1}-Y'_{1,n-1} \right)+O(h^5)
\label{eq_Y0}
\end{align}

Let us now use Eq.~\eqref{eq_taylor_minus} with $\alpha=1$, $q=1$, reuse Eq.~\eqref{eq_taylor_plus} with $\alpha=1$, $q=2$ and eliminate $Y^{(4)}_{1,n}$:
\begin{align}
Y'_{1,n+1}-Y'_{1,n-1}=2hY_{1,n}''+\frac{h}{3}\left( Y''_{1,n+1}+Y''_{1,n-1}-2Y''_{1,n} \right)+O(h^5)
\label{eq_dY1}
\end{align}
Using Eq.~\eqref{eq_dY1} in Eq.~\eqref{eq_Y0}, and using again Eq.~\eqref{eq_amp_lin_cano1} to express $Y''_1(r)$, we get:
\begin{align}
Y_{0,n+1}\left(1+\frac{h^3}{9}\omega'_{n+1}\right)=&\left(2h-\frac{8h^3}{9}\omega_n\right)Y_{1,n}-\frac{4h^3}{9}\omega'_n Y_{0,n}
-\frac{2h^3}{9}\omega_{n+1} Y_{1,n+1}
\nonumber\\
&
-\frac{2h^3}{9}\omega_{n-1} Y_{1,n-1}+\left(1-\frac{h^3}{9}\omega'_{n-1}\right) Y_{0,n-1}+O(h^5)
\label{eq_scheme_Y0}
\end{align}

Finally, we use the latter equation to eliminate $Y_{0,n+1}$ in Eq.~\eqref{eq_scheme_Y1_Numerov}:
\begin{align}
&Y_{1,n+1}\left[1+\frac{h^2}{3}\omega_{n+1}-\left(\frac{h^5}{27}\omega'_{n+1}\omega_{n+1}\right)\left(1+\frac{h^3}{9}\omega'_{n+1}\right)^{-1}\right]\nonumber\\
&=\left(-\frac{h^2}{6}\omega'_{n+1}\right)\left(1+\frac{h^3}{9}\omega'_{n+1}\right)^{-1}\nonumber\\
&\hphantom{=}\cdot
\left[ \left(2h-\frac{8h^3}{9}\omega_n\right)Y_{1,n}-\frac{4h^3}{9}\omega'_n Y_{0,n}
-\frac{2h^3}{9}\omega_{n-1} Y_{1,n-1}+\left(1-\frac{h^3}{9}\omega'_{n-1}\right) Y_{0,n-1}\right]\nonumber\\
&\hphantom{=}+Y_{1,n}\left(2-\frac{10h^2}{3}\omega_{n}\right)
+Y_{0,n}\left(-\frac{10h^2}{6}\omega'_{n}\right)
+Y_{1,n-1}\left(-1-\frac{h^2}{3}\omega_{n-1}\right)
+Y_{0,n-1}\left(-\frac{h^2}{6}\omega'_{n-1}\right)
\nonumber\\
&\hphantom{=}+O(h^6)
\label{eq_scheme_Y1}
\end{align}
Thus, we get a fully explicit expression for $Y_{1,n+1}$, resorting only to values of $Y_1$ and $Y_0$ at the points $n$ and $n-1$. At each step, we use Eq.~\eqref{eq_scheme_Y1} to compute $Y_{1,n+1}$ and then Eq.~\eqref{eq_scheme_Y0} to compute $Y_{0,n+1}$. In practice, we use this numerical scheme in the inward direction. This numerical scheme is of order 6 for $Y_1$ (regular Numerov scheme), and of order 5 for $Y_0$, which is the main quantity of interest. This is to be compared with order 8 of HPC scheme used by Bar Shalom and coauthors. 

Finally, looking at Eq.~\eqref{eq_scheme_Y1}, we can see that error amplification is to be expected whenever $h^2\omega_{k,\ell,n}\gtrsim 1$ or $h^2\omega_{k,\ell,n}'\gtrsim 1$, in the same way as pointed out in \cite{BarShalom96}. Even though in our case, the numerical scheme is much less sensitive to the accumulation of numerical errors, we still need a complementary numerical scheme for coarse grids, or for the outermost region whenever a grid with increasing spacing is used. A simple solution is to adapt the modified predictor-corrector scheme proposed in \cite{BarShalom96}.

\subsection{Outermost region}
For coarse grids, or in the outermost region of a grid of increasingly spaced points, where $h^2\omega_{k,\ell}(r)>1$, we propose an adaptation of the modified predictor corrector of Bar Shalom \textit{et al.}. The key steps are 1) using the slow variation of the amplitude to build a predictor for $Y_1$, $Y_{1}''$
2) using Eq.~\eqref{eq_amp_lin_cano1} to obtain $Y_0$ from $Y_1''$, $Y_1$, thus avoiding error amplification. 

Using Eq.~\eqref{eq_taylor_minus} with $\alpha=1$, $k=0$, we have
\begin{align}
Y_{1,n+1}-Y_{1,n-1}=2Y_{1,n}+h^2Y''_{1,n}+O(h^4)
\label{eq_mod_pred_eq1}
\end{align}
Let us assume that $Y''_{1}$ is varying slowly and thus is roughly equal to its average value over three successive grid point:
\begin{align}
Y''_{1,n}\approx \frac{1}{3}\left(Y''_{1,n-1}+Y''_{1,n}+Y''_{1,n+1}\right)
\label{eq_approx_slow_var_predict}
\end{align}
Using this approximation in Eq.~\eqref{eq_mod_pred_eq1}, one obtains:
\begin{align}
Z_{1,n+1}+Z_{1,n-1}=3Y_{1,n}-Z_{1,n}+O(h^4)
\label{eq_mod_pred_eq2}
\end{align}
where we define $Z_{1}$ as follows:
\begin{align}
Z_{1,n}=Y_{1,n}-\frac{h^2}{3}Y''_{1,n}
\end{align}
Equation~\eqref{eq_mod_pred_eq2} may be used as a predictor for $Z_1$:
\begin{align}
Z_{1,n+1}^\text{p}=3Y_{1,n}-Z_{1,n}-Z_{1,n-1}+O(h^4)
\label{eq_pred_mod_pred_corr}
\end{align}
Let us now use Eq.~\eqref{eq_taylor_minus} with $\alpha=0$, $q=0$:
\begin{align}
Y_{0,n+1}-Y_{0,n-1}=2hY_{0,n}'+\frac{h^3}{3}Y^{(3)}_{0,n}+O(h^5)
\end{align}
Using Eq.~\eqref{eq_amp_lin_cano0}, and the definition of $Z_{1}$, we obtain:
\begin{align}
Y_{0,n+1}-Y_{0,n-1}=h\left(3Y_{1,n}-Z_{1,n}\right)+O(h^5)
\label{eq_mod_pc_Y0}
\end{align}
We finally use Eq.~\eqref{eq_amp_lin_cano1} in order to obtain $Y_1$ from $Y_1''$:
\begin{align}
&Y_{1,n+1}=\frac{-1}{4\omega_{k,\ell,n+1}}\left(Y_{1,n+1}''+2\omega_{k,\ell,n+1}'Y_{0,n+1}\right)\\
&Y_{1,n+1}=\frac{-1}{4\omega_{k,\ell,n+1}}\left[\frac{3}{h^2}\left(Y_{1,n+1}-Z_{1,n+1}\right)+2\omega_{k,\ell,n+1}'\left(Y_{0,n+1}+h(3Y_{1,n}-Z_{1,n})\right)\right]
\end{align}
Because Eq.~\eqref{eq_amp_lin_cano1} is linear, we obtain a linear algebraic equation for $Y_{1,n+1}$, which we readily solve analytically:
\begin{align}
Y_{1,n+1}=\left(1+\frac{3}{4h^2\omega_{k,\ell,n+1}}\right)^{-1}
\left[\frac{3}{4h^2\omega_{k,\ell,n+1}}Z_{1,n+1}-2\omega_{k,\ell,n+1}'\left(h(3Y_{1,n}-Z_{1,n})+Y_{0,n-1}\right)\right]
\label{eq_mod_pc_Y1}
\end{align}
In the method of \cite{BarShalom96}, the equation was nonlinear and one step of a Newton method was used. When we use Eq.~\eqref{eq_mod_pc_Y1}, the value of $Z_{1,n+1}$ is estimated using the predictor $Z_{1,n+1}^\text{p}$ of Eq.~\eqref{eq_pred_mod_pred_corr}. Looking at Eq.~\eqref{eq_mod_pc_Y1}, we clearly see that an amplification of the error on $Z_{1,n+1}^\text{p}$ will occur when $4h^2\omega_{k,\ell,n}/3\lesssim 1$.

\subsection{Method summary}

\begin{table}[h]
\begin{tabular}{c|c|c}
Innermost region &	Intermediate region 		&  Outermost region \\
$[r_\text{min},r_\text{match}]$ & $[r_\text{match},r_\text{switch}]$ & $[r_\text{switch}, r_\text{max}]$ \\
\hline
$h^2\omega_{k,\ell}(r)<1$ & $h^2\omega_{k,\ell}(r)<1$ & $h^2\omega_{k,\ell}(r)>1$\\
Numerov scheme for $R(r)$ &
Eqs~\eqref{eq_scheme_Y1} and \eqref{eq_scheme_Y0}&
Eqs~\eqref{eq_pred_mod_pred_corr}, \eqref{eq_mod_pc_Y0} and \eqref{eq_mod_pc_Y1}\\
Outwards & Inwards & Inwards
\end{tabular}
\caption{Summary of the numerical method proposed in the present work.\label{tab_regions}}
\end{table}

To summarize, our numerical methods is as follows (see also Table~\ref{tab_regions}). From zero to a matching point, we use the usual Numerov scheme outwards, in order to solve Eq.~\eqref{eq_Schrodinger} for $R(r)$. The inner boundary condition that we use corresponds to a radial wavefunction regular at zero. The chosen matching point $r_\text{match}$ is either located near the first maximum of $R(r)$, or at the point where some fraction of the sampling limit is reached, depending on which is closest to zero. 

We then use Eqs~\eqref{eq_pred_mod_pred_corr}, \eqref{eq_mod_pc_Y0} and \eqref{eq_mod_pc_Y1} to propagate the solution inwards, starting from the outer boundary condition, until reaching a point where $h^2\omega_{n}\lesssim 1$. The outer boundary condition is given by Eq.~\eqref{MPA_def_A}, with $R$, $Q$ given by Eqs~(\ref{eq_R_Bessel_Coulomb},\ref{eq_Q_Bessel_Coulomb}). 

In the region where $h^2\omega_{n}\lesssim 1$, we use Eqs.~\eqref{eq_scheme_Y1} and \eqref{eq_scheme_Y0} to propagate the solution further inwards, up to the matching point. 

Matching is performed in order to normalize the inner part of $R(r)$ according to the outer boundary condition, and to determine the phase at the matching radius. The phase function is subsequently obtained by integrating outwards the function $\phi(r)-kr$ using Simpson's rule. 
%It is also fully possible to take advantage of the fact that our method also give access to the derivative $Y_1(r)$.

\section{Test of the method}
We tested the present method using linear grids, quadratic grids, exponentially spaced grids, and mixed linear-exponential grids. The relations between the radius $\rho$ in these grids, and the parameter $r$ which is uniformly sampled are recalled in Table~\ref{tab_grids}. Together with the change of variable from $\rho$ to $r$, we perform the change of function $A(\rho)\rightarrow \hat{A}(r)$, which leaves Eqs~(\ref{eq_Schrodinger},\ref{MPA_eq_amp},\ref{eq_amp_lin}) invariant in form:
\begin{align}
\hat{A}(\hat{r}(\rho))=A(\rho)/f(\rho)
\label{eq_grid_change_funct}
\end{align}
The relevant functions $f(\rho)$ for each grid are also given in Table~\ref{tab_grids}.

\begin{table}[b]
\begin{tabular}{c|c|c}
Grid type			& Change of variable 	& Function $f$ of Eq.~\eqref{eq_grid_change_funct} \\
\hline
Linear				& $r=\rho$					& $1$		\\
\hline
Quadratic			& $r=\sqrt{\rho}$  			& $\rho^{1/4}$	\\
\hline
Exponential			& $r=\ln(\rho)$				& $\sqrt{\rho}$\\
\hline
Exponential-linear	& $r=\ln(\rho)+a\,\rho$		& $\sqrt{\dfrac{\rho}{1+a\rho}}$\\
\end{tabular}
\caption{Change of variable and function $f$ corresponding to various grids.\label{tab_grids}}
\end{table}	

\begin{figure}[t]
\centerline{
\includegraphics[width=8cm]{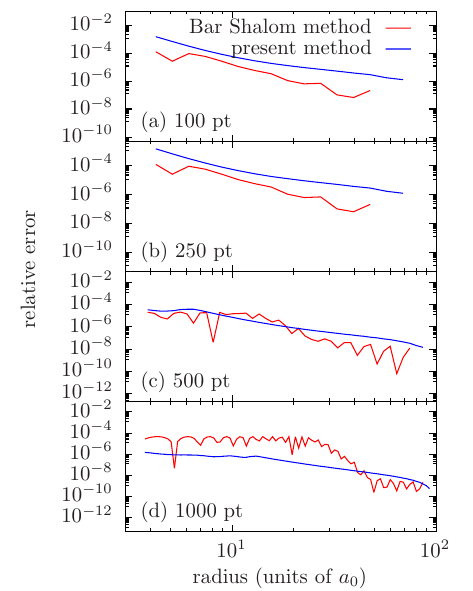}
\includegraphics[width=8cm]{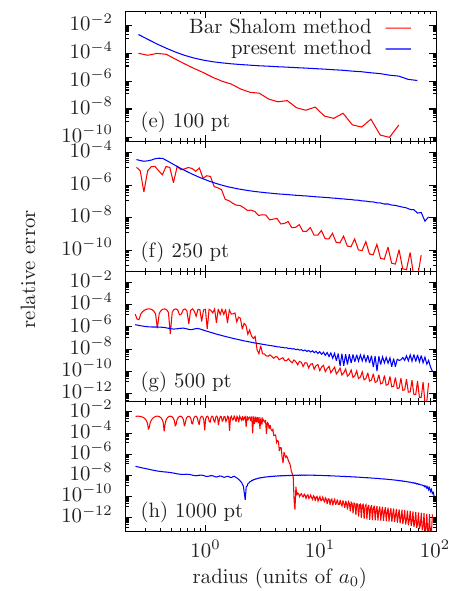}
}
\caption{Relative error on the amplitude function obtained with Bar Shalom method and with the present method. Case of zero potential, for $\ell=2$, $k=1\,a_0^{-1}$(a-d) and $k=15\,a_0^{-1}$ (e-h) using a exponentially spaced grid spanning from $10^{-6}$ to $10^2$\,$a_0$ with $100$ (a,e), $250$ (b,f), $500$ (c,g), and $1000$ (d,h) points, respectively. The error is showed only from the matching radius to the grid outer boundary.
\label{fig_error_log_grid_k_1_15}}
\end{figure}

Figure~\ref{fig_error_log_grid_k_1_15} presents the relative error on the amplitude function obtained with our method, compared to that of \cite{BarShalom96}, in the case of zero potential, for $\ell=2$, $k=1\,a_0^{-1}$ and $15\,a_0^{-1}$. In order to remain as close as possible to Reference \cite{BarShalom96}, we perform this comparison using an exponentially spaced grid, varying the number of grid points.

Errors in the outermost region seem systematically larger with the present method than with the method of \cite{BarShalom96}. This may indicate that the approximation of Eq.~\eqref{eq_approx_slow_var_predict} is less justified for $Y^{(3)}(r)$ than it is for $A''(r)$. However, the largest contribution to the numerical errors stems from error accumulation in the intermediate region, where the amplitude is varying more significantly. For that reason, we do not consider the numerical scheme used in the outermost region as a limiting factor for our method. 

Using a coarse grid (see Fig.~\ref{fig_error_log_grid_k_1_15}\,a and e), for which the method of \cite{BarShalom96} performs well, error growth in the intermediate region is stronger with our fifth-order scheme than with the HPC scheme, which is of order 8. This is fully expected. 

Considering finer grids (see Fig.~\ref{fig_error_log_grid_k_1_15}\,b--d and f--h), one clearly sees the effect of the spurious component that appears with the method of \cite{BarShalom96}, which puts a lower bound on the precision that may be achieved, and ultimately ruins the solution. On the contrary, the present method exhibits a continuous improvement of precision, when one refines the grid. Beyond some grid refinement, it may become relevant to improve the solution in the outermost region. This can be done as suggested in \cite{BarShalom96}, by making some iterative steps after applying the modified predictor-corrector scheme.

%The same behavior was observed with other types of grid: linear, quadratic, and mixed linear-exponential grids. We present in Fig.~xx various examples.

\begin{figure}[h]
\centerline{
\includegraphics[width=8cm]{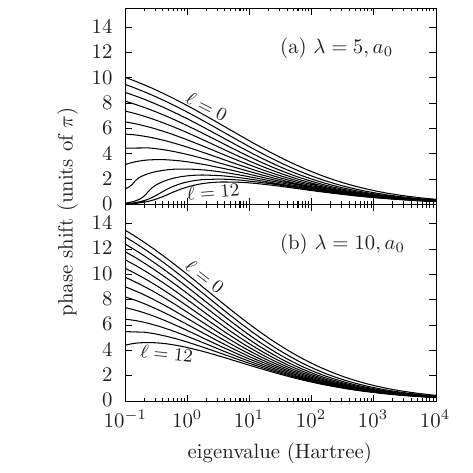}
}
\caption{Scattering phase-shifts calculated using the present method, for two cases of fully screened Coulomb potential $-Z e^{-r/\lambda}/r$ with $Z=26$, $\lambda=5\,a_0$ (a) and $10\,a_0$ (b) for eigenvalues $k^2/2$ ranging from $10^{-1}$ to $10^{4}$ Hartree and orbital quantum number $\ell$ ranging from $0$ to $12$.
\label{fig_phase_shift}}
\end{figure}

As an example of systematic application of the present numerical method, we present on Fig~\ref{fig_phase_shift} the phase-shifts as functions of eigenvalue (i.e. orbital energy) for two fully screened Coulomb potentials with $Z=26$ and decay length $\lambda=5$ and $10\,a_0$, for orbital number $\ell$ ranging from $0$ to $12$. These curves were obtained using the present method on an exponentially spaced grid of $1000$ points, spanning from $10^{-6}\,a_0$ to $100\,a_0$. Using the Milne PA representation with the present numerical method allows one to compute phase shift for arbitrarily high eigenvalues keeping a relatively coarse radial grid.

\section{Conclusion}
In the present article, we first recall the basics of Milne's phase-amplitude representation. We discuss the limitations of some known methods to solve Milne's amplitude equation, including the fully-explicit predictor-corrector method of Bar Shalom et al. and adaptive grid refinement methods, both explicit and implicit. A fully explicit numerical method based on the linear equation derived by Kiyokawa is proposed as a solution to address these limitations. The key point being that, solving a linear equation, one gets rid of any coupling to the spurious, rapidly varying solution of the amplitude equation and obtains a much more robust method. The method is tested and its performance is checked against that of Bar Shalom et al.. It is shown that, while being of lower order than the predictor-corrector method used in the work of Bar Shalom et al., the present method yields a systematic improvement of the precision when refining the grid. Upcoming work will consider the applications of this method to the calculation of atomic cross-sections.

\section*{Aknowdedgements}
The authors' respective contributions are as follows. RP studied the limitations of the method of Bar Shalom et al., conceptualized the present study, proposed the explicit numerical method and performed the comparisons with that of Bar Shalom et al.. MT explored the use of the Runge-Kutta-Merson method, and performed the comparisons with the Kaps-Rentrop method.
RP would like to acknowledge discussions with B. G. Wilson and M. Klapisch on this subject at the RPHDM conferences in 2008 and 2010. 
%MT would like to acknowledge Pautard for suggesting the Runge-Kutta-Merson method as a candidate numerical method.

\bibliographystyle{unsrt}
\bibliography{bibexport.bib}

\begin{thebibliography}{10}

\bibitem{Kiyokawa15}
Shuji Kiyokawa.
\newblock {Exact solution to the Coulomb wave using the linearized
  phase-amplitude method}.
\newblock {\em AIP Advances}, 5:087150, 2015.

\bibitem{Blenski13}
T.~Blenski, R.~Piron, C.~Caizergues, and B.~Cichocki.
\newblock {Models of atoms in plasmas based on common formalism for bound and
  free electrons}.
\newblock {\em High Energy Density Physics}, 9:687–695, 2013.

\bibitem{Piron24}
R.~Piron.
\newblock {Atomic Models of Dense Plasmas, Applications, and Current
  Challenges}.
\newblock {\em Atoms}, 12:26, 2024.

\bibitem{Liberman79}
D.~A. Liberman.
\newblock {Self-consistent field model for condensed matter}.
\newblock {\em Phys. Rev. B}, 20(12):4981–4989, 1979.

\bibitem{Wilson06}
B.~Wilson, V.~Sonnad, P.~Sterne, and W.~Isaacs.
\newblock {\textsc{Purgatorio} - a new implementation of the \textsc{Inferno}
  algorithm}.
\newblock {\em J. Quant. Spectrosc. Radiat. Transf.}, 99:658–679, 2006.

\bibitem{Piron11}
R.~Piron and T.~Blenski.
\newblock {Variational-average-atom-in-quantum-plasmas (VAAQP) code and virial
  theorem: Equation-of-state and shock-Hugoniot calculations for warm dense Al,
  Fe, Cu, and Pb}.
\newblock {\em Phys. Rev. E}, 83:026403, 2011.

\bibitem{Piron18}
R.~Piron and T.~Blenski.
\newblock {Average-atom model calculations of dense-plasma opacities: Review
  and potential applications to white-dwarf stars}.
\newblock {\em Contributions to Plasma Physics}, 58:30–41, 2018.

\bibitem{Shannon49}
C.E. Shannon.
\newblock {Communication in the Presence of Noise}.
\newblock {\em Proceedings of the IRE}, 37:10–21, 1949.

\bibitem{Starrett23}
C.~E. Starrett and N.~R. Shaffer.
\newblock {Average-atom model with Siegert states}.
\newblock {\em Phys. Rev. E}, 107:025204, Feb 2023.

\bibitem{Batishev07}
Pavel~A. Batishchev and Oleg~I. Tolstikhin.
\newblock {Siegert pseudostate formulation of scattering theory: Nonzero
  angular momenta in the one-channel case}.
\newblock {\em Phys. Rev. A}, 75:062704, Jun 2007.

\bibitem{Milne30}
W.~E. Milne.
\newblock {The numerical determination of characteristic numbers}.
\newblock {\em Phys. Rev.}, 35:863–867, 1930.

\bibitem{Goldstein58}
M.~Goldstein and R.~M. Thaler.
\newblock {Bessel Functions for Large Arguments}.
\newblock {\em Mathematical Tables and Other Aids to Computation}, 12:18–26,
  1958.

\bibitem{BarShalom96}
A.~Bar-Shalom, M.~Klapisch, and J.~Oreg.
\newblock {Phase-amplitude algorithms for atomic continuum orbitals and radial
  integrals}.
\newblock {\em Comput. Phys. Commun.}, 93:21–32, 1996.

\bibitem{Wilson03}
B.~Wilson, V.~Sonnad, P.~Sterne, and W.~Isaacs.
\newblock {Improvements in the phase-amplitude method for calculating free-free
  gaunt factors and spherical bessel function of high angular momentum}.
\newblock {\em J. Quant. Spectrosc. Radiat. Transf.}, 81:499–512, 2003.

\bibitem{Seaton62}
M~J Seaton and G~Peach.
\newblock {The Determination of Phases of Wave Functions}.
\newblock {\em Proceedings of the Physical Society}, 79(6):1296, jun 1962.

\bibitem{Trafton71}
L~Trafton.
\newblock {A rapid numerical solution to the radial Schroedinger equation in
  the oscillatory region}.
\newblock {\em Journal of Computational Physics}, 8:64–72, 1971.

\bibitem{Rawitscher15}
George Rawitscher.
\newblock {A spectral Phase–Amplitude method for propagating a wave function
  to large distances}.
\newblock {\em Computer Physics Communications}, 191:33–42, 2015.

\bibitem{BarShalom97}
A.~Bar-Shalom, J.~Oreg, and M.~Klapisch.
\newblock {Non-LTE Superconfiguration Collisional Radiative Model}.
\newblock {\em J. Quant. Spectrosc. Radiat. Transf.}, 58:427–439, 1997.

\bibitem{BarShalom01}
A.~Bar-Shalom, M.~Klapisch, and J.~Oreg.
\newblock {HULLAC, an integrated computer package for atomic processes in
  plasmas}.
\newblock {\em J. Quant. Spectrosc. Radiat. Transf.}, 71:169–188, 2001.

\bibitem{PironPhD}
R.~Piron.
\newblock {\em {Variational Average-Atom in Quantum Plasmas (VAAQP)}}.
\newblock PhD thesis, École Polytechnique, 2009.

\bibitem{Good53}
R.~H. Good.
\newblock {The Generalization of the WKB Method to Radial Wave Equations}.
\newblock {\em Phys. Rev.}, 90:131–137, 1953.

\bibitem{Rosen64}
M.~Rosen and D.~Yennie.
\newblock {A Modified WKB Approximation for Phase Shifts}.
\newblock {\em Journal of Mathematical Physics}, 5:1505–1515, 1964.

\bibitem{Wald74}
S.~S. Wald and P.~Lu.
\newblock {Modofied WKB approximation for phase shifts of an attractive
  singular potential}.
\newblock {\em Phys. Rev.}, 10:3434–3440, 1974.

\bibitem{Stein87}
J.~Stein, Akiva Ron, I.~B. Goldberg, and R.~H. Pratt.
\newblock {Generalized WKB approximation to nonrelativistic normalizations and
  phase shifts in a screened Coulomb potential}.
\newblock {\em Phys. Rev. A}, 36:5523–5529, 1987.

\bibitem{Calogero63}
F.~Calogero.
\newblock {A novel approach to elementary scattering theory}.
\newblock {\em Il Nuovo Cimento}, 27:261–302, 1963.

\bibitem{Green82}
A.~E.~S. Green, D.~E. Rio, P.~F. Schippnick, J.~M. Schwartz, and P.~S. Ganas.
\newblock {Analytic Phase Shifts for Yukawa Potentials}.
\newblock {\em International Journal of Quantum Chemistry}, 16:331–343, 1982.

\bibitem{Blenski88}
T.~Blenski and J.~Ligou.
\newblock {An improved shooting method for one-dimensional schrödinger
  equation}.
\newblock {\em Comput. Phys. Commun.}, 50:303–311, 1988.

\bibitem{Nikiforov}
A.~F. Nikiforov, V.~G. Novikov, and V.~B. Uvarov.
\newblock {\em {Quantum-Statistical Models of Hot Dense Matter}}.
\newblock Birkhauser, 2005.

\bibitem{AbramowitzStegun}
M.~Abramowitz and I.~A. Stegun.
\newblock {\em {Handbook of Mathematical Functions with Formulas, Graphs, and
  Mathematical Tables.djvu}}.
\newblock Dover Publication, Inc, New York, 1965.

\bibitem{Hamming}
R.~W. Hamming.
\newblock {\em {Numerical Methods for Scientists and Engineers}}.
\newblock Dover Publication, Inc, 1986.

\bibitem{Kaps79}
P.~Kaps and P.~Rentrop.
\newblock {Generalized Runge-Kutta Methods of Order Four with Stepsize Control
  for Stiff Ordinary Differential Equations}.
\newblock {\em Numerische Mathematik}, 33:55–68, 1979.

\bibitem{NumericalRecipes}
W.~H. Press, S.~A. Teukolsky, W.~T. Vetterling, and B.~P. Flannery.
\newblock {\em {Numerical Recipes, The Art of Scientific Computing}}.
\newblock Cambridge University Press, 2007.

\bibitem{Merson57}
R.~H. Merson.
\newblock {An operational method for the study of integration processes}.
\newblock {\em Proc. Symp. Data Processing, Weapons Res. Establ. Salisbury},
  page 110–125, 1957.

\end{thebibliography}

\end{document}